\documentclass[aps,prl,floatfix,twocolumn,showpacs]{revtex4}
\usepackage{epsfig}
\usepackage{amsmath}
\begin{document}

\hsize\textwidth\columnwidth\hsize\csname@twocolumnfalse\endcsname

\title{Entanglement distillation by adiabatic passage in coupled quantum dots}

\author{Jaroslav Fabian and Ulrich Hohenester}
\affiliation{Institute of Physics, Karl-Franzens
University, Universit\"{a}tsplatz 5, 8010 Graz, Austria}

\vskip1.5truecm
\begin{abstract}
Adiabatic passage of two correlated electrons in three coupled quantum dots is shown
to provide a robust and controlled way of distilling, transporting and detecting
spin entanglement, as well as of measuring the rate of spin disentanglement.
Employing tunable interdot coupling the scheme creates,
from an unentangled two-electron state, a superposition of 
spatially separated singlet and triplet states. 
A single measurement of a dot population (charge) collapses
the wave function to either of these states, realizing
entanglement to charge conversion. The scheme is robust, with the efficiency close to 100\%,
for a large range of realistic spectral parameters.
\end{abstract}
\pacs{03.67.Mn, 03.67.Hk, 03.67.Lx, 73.63.Kv}
\maketitle

Creation and detection of spin entanglement is a major task 
for quantum information processing\cite{Bennett2000:N}. 
A particular implementation of the processing relies on 
electron spins in coupled quantum dots, 
proposed as qubits for quantum inverters \cite{Bandyopadhyay1997:SM}
and for universal gating in quantum computation \cite{Loss1998:PRA}.
It has been proposed that entangled two-electron spin states 
in quantum dots can be produced by tuned quantum
gates \cite{Loss1998:PRA, Burkard1999:PRB, Hu2000:PRA}, by filtering
through time-dependent barriers \cite{Hu2004:PRB}, or by projective
measurements \cite{Ruskov2003:PRB, Stace2004:P}. Entanglement 
is proposed to be detected by current noise measurements \cite{Loss2000:PRL}.
Impressive recent progress in coherent control of electronic states
in quantum dots \cite{vanderWiel2003:RMP,Hayashi2003:PRL, Petta2004:PRL, Petta2004:P} 
and spin coherence \cite{Hanson2003:PRL} gives strong impetus to these
concepts.

Here we introduce a scheme for performing spin entanglement distillation.
The remarkable feature of the scheme is that, unlike previous proposals,
it is also capable of entanglement detection, transport, as well as disentanglement
measurement, all in a robust way, without the need for fine tuning 
or precise knowledge of spectral or pulse parameters. The scheme is 
based on our finding of a strong correlation between adiabatic passage and entanglement: 
a single adiabatic pulse induces entirely different adiabatic passages
of different Bell (maximally entangled) spin states. 
We call our scheme, which can be realized by current experimental techniques,
entanglement distillation by adiabatic passage (EDAP).

We demonstrate the scheme on two electrons
in three coupled quantum dots. Starting from an unentangled pair,
a combination of temporal pulse sequences of interdot couplings 
spatially separates singlet and triplet states. Entanglement
is converted to charge whose detection uniquely gives  a triplet 
or a singlet. The same
principle is used for entanglement detection. The scheme transports
triplets only, leaving singlets, allowing for
selective transport of entangled pairs. 
Our work is motivated by quantum control techniques for atomic systems
such as
adiabatic passage and stimulated Raman adiabatic passage (STIRAP)
\cite{Bergmann1998:RMP} which are increasingly used in solid state physics
\cite{Hohenester2004:P, Brandes2004:P}. For example, STIRAP has 
been recently proposed to transport single electrons 
\cite{greentree2004:P}, and adiabatic passage to spatially separate
electron singlet pairs\cite{Zhang2004:PRB}, in coupled dots. 

\begin{figure}
\centerline{\psfig{file=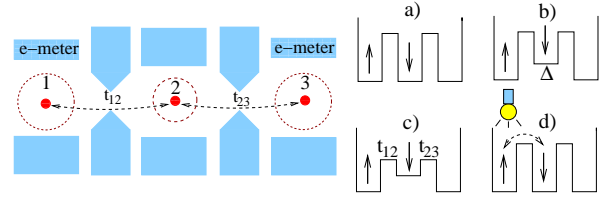,width=0.9\linewidth}}
\caption{Entanglement distillation by adiabatic passage. 
Three quantum dots are coupled via electrode-defined barriers giving tunnel couplings
$t_{12}$ and $t_{23}$. The ground state energy of dot $2$ is
shifted by $\Delta$. The charge on dots 1 or 3 is detected by 
electrometers. On the right the four figures show the scheme at work [the
light bulb in d) is an electrometer]
}
\label{fig:1}
\end{figure}

We model the physics of two electrons in three coupled dots 
(Fig. \ref{fig:1}) by the time-dependent Hubbard Hamiltonian
\begin{equation} \label{eq:Hubbard}
H= \sum_{i\lambda} \varepsilon_i n_{i\lambda} +
  \sum_{i<j,\lambda\lambda'} U_{i\lambda,j\lambda'} n_{i\lambda}n_{j\lambda'}  +  
\sum_{ij, \lambda} t_{ij} a_{i\lambda}^+a_{j\lambda},
\end{equation}
with the Fermi creation ($a^+_{i\lambda}$) and annihilation ($a_{i\lambda}$) operators
for dot $i$ (1, 2, and 3) and spin $\lambda=\uparrow, \downarrow$,
and number operators $n_{i\lambda}=a_{i\lambda}^+a_{i\lambda}$. 
The confining energies $\varepsilon_{i}$ do not depend on spin. 
We take $\varepsilon_1=\varepsilon_3=0$, while
setting an offset for the middle dot $\varepsilon_2=\Delta$. 
The offset can be controlled electrostatically,
or it can be fixed within a useful spectral range as shown below.  
We take the on-site Coulomb repulsion $U_{i\uparrow, i\downarrow}=U$ 
to be the same for all dots; similarly for the off-site interactions
$U_{i,\lambda;i+1,\lambda'}=V$, and zero otherwise. 
Hopping integrals representing interdot couplings are 
$t_{ij}$. For our system only $t_{12}$ and $t_{23}$ are not zero and
depend on time $t$, so that $H=H(t)$. 
The interdot couplings are modulated by electrostatic gates 
defining interdot barriers. 
The spectral scales are meVs, with $t \ll U$ for realistic systems. 
In the examples below we use generic values of 
$U=1$ meV, $V=0$ or 0.1 meV, and maximum hoppings smaller than 0.1 meV. 
Precise values will not be relevant.

\begin{figure}
\centerline{\psfig{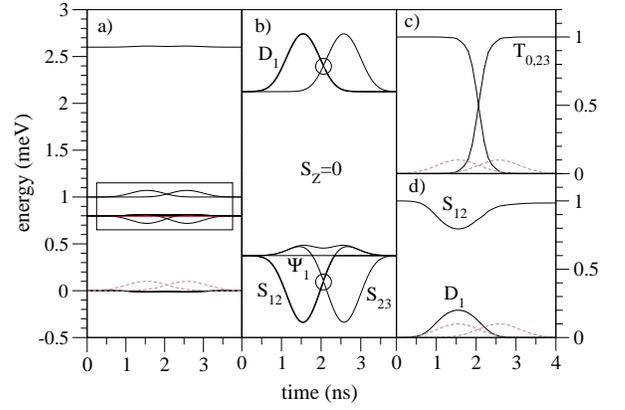}}
\caption{ (a) Temporal evolution of the two-electron spectrum (solid lines)
of Hamiltonian $H$ in the presence of two overlapping Gaussian
pulses (dashed) of $t_{12}$ and $t_{23}$. 
The spectra are plotted for $U=1$ meV, $V=0$, and $\Delta=0.8$ meV. The pulses
of $t_{12}(t)$ and $t_{23}(t)$ have widths  $\tau \approx 500 $ ps.
(b) States with $S_z=0$ relevant
for EDAP, from the box in a). There is a level
repulsion (anticrossing) inside the circles, where the passage is rapid. At other
two crossings there is no repulsion. The horizontal
line is  the trapped state $\Psi_1$. 
(c) Counterintuitive passage scheme for $\Psi_{1}$ showing probabilities $p$ of
finding states $|T_0\rangle_{12}$ and $|T_0\rangle_{23}$. (d)
Passage scheme for $|S\rangle_{12}$ showing the probabilities $p$
of observing $|S\rangle_{12}$ and $|D_1\rangle$.
}
\label{fig:2}
\end{figure}

The time dependent spectrum of $H$, in the presence of interdot
coupling pulses, is shown in Fig. \ref{fig:2}a. We take Gaussian
pulses of the form $t_{ij}(t)=t_0\exp(t^2/2\tau^2)$, where $t_0$ is
the maximum pulse strength and $\tau$ is the dispersion. 
The overlap between the pulses is taken to 
be $2\tau$, the width of one pulse. There are three
weakly coupled groups of states. The lowest states
with energy $E\approx 0$ consist of electrons occupying mainly
dots 1 and 3. The highest state, of $E\approx U+2\Delta$,
is for a double occupancy of dot 2. The states relevant for EDAP
have $E\approx U, \Delta$, and comprise electron singlets and
triplets on neighboring dots. These states are magnified
in Fig. \ref{fig:2}b. To simplify notation we introduce the following labels
for triplet $T$ and singlet $S$ states 
on dots $i$ and $j$ (assuming $i<j$), as well as for double occupancy 
states $D$:
\begin{eqnarray}
|T_{1} \rangle_{ij} &=& a^+_{i\uparrow}a^+_{j\uparrow} |0\rangle, \,\,\,\,\,
|T_{-1} \rangle_{ij}= a^+_{i\downarrow}a^+_{j\downarrow} |0\rangle, \\
|T_0\rangle_{ij}& = & (1/\sqrt{2})(a^+_{i\uparrow}a^+_{j\downarrow} +
a^+_{j\uparrow}a^+_{i\downarrow})|0\rangle, \\
|S\rangle_{ij}& = & (1/\sqrt{2}) a^+_{i\uparrow}a^+_{j\downarrow} -
a^+_{j\uparrow}a^+_{i\downarrow}|0\rangle, \\
|D\rangle_{i}&=&a^+_{i\uparrow}a^+_{i\downarrow}|0\rangle.
\end{eqnarray}
Here $|0\rangle$ is the vacuum. The triplet states $|T_{S_z}\rangle$ 
are labeled by their spin $S_z$. States $|T_0\rangle$ and 
$|S\rangle$ are spin entangled.

We first summarize EDAP steps and then discuss the physics in
detail. The scheme is shown in Fig. \ref{fig:1}: (a) Start with two
uncoupled electrons occupying neighboring dots 1 and 2.
(b) Raise slowly the energy of the middle
dot 2 to $\Delta$ being on the scale of $U$ (this step is not
necessary if $\Delta$ is built in). (c) Apply an overlapping
pulse sequence of $t_{12}$ and $t_{23}$ (order not relevant).
After the pulses fade away, $\Delta$ can be 
switched back to zero, if necessary. The resulting 
state is with a high probability a superposition of a singlet
state, spread over dots 1 and 2, and triplet states,  on dots 2 and
3. A detection of (the absence of) charge in dot 1, collapses the wave 
function to the singlet (triplet). Mathematically, an initial
two-electron state $\Psi(t=0)$ localized on dots
1 and 2, is a superposition
\begin{equation} \label{eq:super}
\Psi(0)= a|S\rangle_{12}+b|T_0\rangle_{12} + c|T_{1}\rangle_{12} + d |T_{-1}\rangle_{12}.
\end{equation}
After EDAP, the state will be
\begin{equation}
\Psi(\infty)= a'|S\rangle_{12}+b'|T_0\rangle_{23} + c'|T_{1}\rangle_{23} + d' |T_{-1}\rangle_{23},
\end{equation}
where the primed coefficients are equal to unprimed up to a phase factor.
The singlet state returns to the initial dots while the triplets are
transported to dots 2 and 3. As a result entanglement is coupled to charge
on dots 1 and 3. The scheme also works as a noninvasive entanglement detector. 
If the
initial state is a singlet, the final state is the same (up to a phase). 
If it is a triplet, the state is shifted in space. Charge measurement on dots
1 or 3, which is a nondemolition measurement for singlet and triplet 
states in the absence of interdot coupling, separates the two.
In general, probabilities of finding, say the singlet in a given initial
state, $|a|^2$ can be obtained by repeating the measurement on the
identically prepared state, detecting a degree of entanglement.
The scheme does not, however, discern the individual triplet states
$|T_0\rangle$ and $|T_{\pm 1}\rangle$
without an additional single-dot control (e.g., spin rotation).
Finally, the scheme detects disentanglement and charge decoherence
by observing systematic deviations from the expected
final states (e.g., detecting charge on {\it both} 1 and 3). 

To demonstrate the scheme we study the evolution
of each of the states in the superposition of Eq. \eqref{eq:super}. 
Consider triplet states first. It is useful
to find the eigenstates of $H$ whose energies do not depend on $t_{12}$
or $t_{23}$; in analogy with quantum optics, we call these
states trapped. There are four two-electron trapped states of $H$: 
\begin{eqnarray}
\Psi_1 &=&\sin\varphi |T_0\rangle_{12}-\cos\varphi|T_0\rangle_{23}, \\
\Psi_2 &=&\sin\varphi |T_1\rangle_{12} -\cos\varphi |T_1\rangle_{23}, \\
\Psi_3 &=&\sin\varphi |T_{-1}\rangle_{12} -\cos\varphi |T_{-1}\rangle_{23}, \\
\Psi_4 &=&\left [ |D\rangle_1 - |D\rangle_2 + |D\rangle_3
\right] / \sqrt{3}.
\end{eqnarray}
The mixing angle $\varphi=\varphi(t)$ is defined by $\tan\varphi=t_{12}/t_{23}$.
States $\Psi_1$ through $\Psi_3$ have energy $V+\Delta$, while
$\Psi_4$, which is trapped only for $\Delta=0$, has energy $U$.
As in STIRAP, which is a technique for population transfer via
trapped states\cite{Bergmann1998:RMP}, states $\Psi_1$ through $\Psi_3$ allow the
passage of an initial triplet state $|T\rangle_{12}$ to 
$|T\rangle_{23}$, or vice versa. Take $\Psi_1$ a an example.  
If the initial state is $|T_0\rangle_{12}$, it 
will be 100\% in $\Psi_1$ for $t_{23}=0$, 
when $t_{12}$ is slowly turned on ($\varphi=\pi/2$). The state is unaltered until
a subsequent overlapping
pulse of $t_{23}$ will smoothly move the state to 
$\Psi_1=|T_0\rangle_{23}$, after $t_{12}$ vanishes ($\varphi=0$). 
During the passage, no state other than the two
triplets is populated. The numerical calculation
is shown in Fig. \ref{fig:2}c, confirming the qualitative
picture. In our context this pulse sequence
($t_{12}$ before $t_{23}$) can be called counterintuitive, 
while the opposite order ($t_{23}$ before $t_{12}$) intuitive.
Transfer through $\Psi_1$ by counterintuitive sequence 
is extremely robust, independent on spectral parameters,
as long as adiabatic conditions, to be specified, hold.  
While adiabatic passage via $\Psi_1$ is a  nine-level
process (there are nine $S_z=0$ basis states), the
scheme with $\Psi_2$ and $\Psi_3$, for transporting
spin unentangled triplets $|T_1\rangle$ and $|T_{-1}\rangle$,
is an exact analogue of the three level STIRAP. 
Triplet states can also be transferred through intuitive
sequencing (not via $\Psi_1$), if $\Delta$ is greater than
the interdot couplings. Such a transfer is less robust,
but for our scheme it is equally satisfactory as
counterintuitive, since we need $\Delta \agt t_{12}, t_{23}$
to transfer singlet states, as shown below.
The fourth trapped state, $\Psi_4$, is a superposition of doubly occupied
states. Because it
cannot be manipulated with interdot couplings, we call this state
globally trapped. We will not use this state below.

Singlet states are not part of the trapped states. If the
initial state is the singlet $|S\rangle_{12}$, the above scheme 
in general leads to an arbitrary superposition of eigenstates of $H$ for
isolated dots. There is, however, a window of energy offsets $\Delta$ where 
the final state will be $|S\rangle_{12}$, up to a phase. Consider states
$|S\rangle_{12}$, $|D\rangle_1$, and $|D\rangle_2$, with average
energies $\Delta+V$, $U$, and $2\Delta+U$, respectively.
If we make $\Delta$ on the same scale as $U$, state $|D\rangle_2$,
as well as all other eigenstates, will not be easily accessible due to 
spectral separation (Fig. \ref{fig:2}a). 
We have an effective two-level system with Hamiltonian 
(up to a constant)
\begin{equation} \label{eq:H}
H'=\frac{1}{2}(\Delta + V -U) \sigma_z + \sqrt{2}t_{12}(t) \sigma_x,
\end{equation}
where $\sigma_\alpha$ are the Pauli matrices. The eigenstates are
\begin{eqnarray}
\Psi_{+}&=& \cos \left (\vartheta/2\right ) |S\rangle_{12} + 
\sin \left (\vartheta/2\right ) |D\rangle_1, \\
\Psi_{-}&=& \sin \left (\vartheta/2\right ) |S\rangle_{12} -
\cos \left (\vartheta/2\right ) |D\rangle_1.
\end{eqnarray}
The mixing angle $\vartheta=\vartheta(t)$, restricted to $[0,\pi]$, is defined by $\tan\vartheta
= 2\sqrt{2}t_{12}/(\Delta + V -U)$. 
The nature of the time evolution of the singlet depends critically on $\Delta$. 
In resonance, $\Delta+V\approx U$, the singlet is initially a superposition
of $\Psi_{+}$ and $\Psi_{-}$. After passage of pulse $t_{12}$ 
the final state will be $ \Psi(\infty)=|S\rangle_{12}\cos\alpha + |D\rangle_1\sin\alpha$,
where the pulse area $\alpha=\int \sqrt{2} t_{12}(t) dt$. By fine tuning the
pulses to $\alpha=\pi$, the final state will be $|S\rangle_{12}$.

The above resonant scheme for singles, though allowing fine control, is not robust: 
it requires both the resonance condition and precise knowledge of the pulse area. 
We instead explore the large spectral window off the resonance. 
For $|\Delta+V-U| \agt t_{12}$, state $|S\rangle_{12}$ will be transported
back to itself, via $\Psi_{+}$. (This is analogous to adiabatically following
of a spin along a magnetic field that rotates along $y$-axis back and forth
adiabatically.) Such a passage is very robust. The two-level picture
is confirmed by the numerical calculation with the full Hamiltonian $H$
in Fig. \ref{fig:2}d.

\begin{figure}
\centerline{\psfig{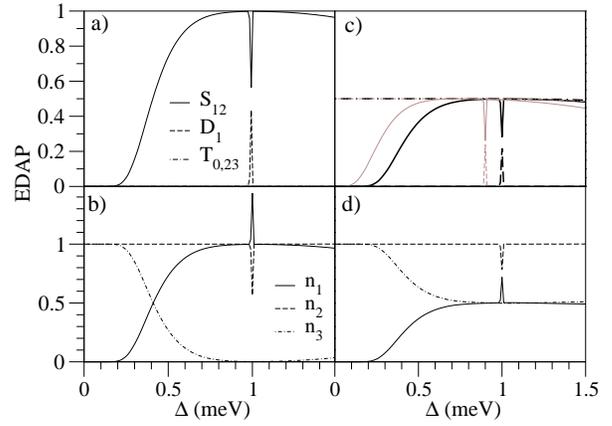}}
\caption{ a) Calculated probabilities and electron populations 
after EDAP passage as a function of $\Delta$, 
with $\Psi(0) = |S\rangle_{12}$ (a and b) and
$\Psi(0) = a_{1\uparrow}^+ a_{2\downarrow}^+|0\rangle$ (c and d).
The thin dotted lines in c) are for $V=0.1$ meV. Pulses are the same as
in Fig. \ref{fig:2}.
}
\label{fig:3}
\end{figure}

Figure \ref{fig:3} shows EDAP results as a function of $\Delta$,
for two initial states. For the selected $\tau$ the initial singlet
returns to the same state at least 90\% of times for $\Delta \agt 0.6$ meV
(unless at resonance visible by spikes). This is closely mirrored
by the charge population of the dots 1 and 3. Dot 2 has always
charge one, except for resonance, in which dot 1 can be doubly occupied.
The influence of off-site Coulomb interaction is seen in Fig. \ref{fig:3}c.
The only effect is shifting the resonance from $\Delta = U$ to $\Delta = U-V$.

What is the condition on the pulse? Passage of state $|l\rangle$ is adiabatic
if $|\langle l | \hbar \dot{H(t)} |k\rangle| \ll (\hbar\omega_{lk})^2$, where
$k$ are other eigenstates of $H(t)$ and $\omega_{lk}$ are the Bohr
frequencies \cite{Messiah:1965}. We give rough estimates for the
limits on pulse dispersion $\tau$ (switching time), based on the
qualitative criterion that the smallest relevant Bohr period needs to
be resolved during the passage.
EDAP comprises four processes:
(i) Adiabatic passage of the triplet state. 
For $|\Delta| \alt t_0$, which can be used for triplet transport, 
this is robust if $\tau \agt \hbar/t_0$. In our scheme 
$\Delta \gg t_0$ and the smallest relevant Bohr energy is $t_0^2/\Delta$. Then
$\tau \agt \tau_L$ where $\tau_L= \hbar/(t_0^2/\tau)$ 
gives the lower limit.
(ii) Adiabatic passage of the singlet. This is a two-level scheme
with states separated by $\sim t_0$. Thus $\tau \agt \hbar/t_0$, 
which is within the range of (ii) and need not be considered extra.
(iii) Rapid passage of the singlet through the anticrossing at time
$t$ given by $\bar{t}=t_{12}(t)=t_{23}(t)$ (Fig. \ref{fig:2}b).
The level repulsion is small, since $|S\rangle_{12}$ couples to
$|S\rangle_{23}$ through spectrally distant states such as
$|S\rangle_{13}$ and $|D\rangle_2$. Interference in the virtual
coupling to these states further reduces the anticrossing. One can
show that the level spacing is $V_g\approx 2 (\bar{t}^2/\Delta)(U-\Delta)/(U+\Delta)$,
vanishingly small at resonance in the oder $O(\bar{t}^2/\Delta)$. 
Rapid passage occurs for $\tau' \alt \hbar/V_g$,
where $\tau'=\tau V_g/t_0$ is the time over which the interdot
coupling changes by $V_g$, relevant for resolving the gap. This
gives $\tau \alt \tau_{U}$ where
$\tau_{U}=\tau_{L}\times (\Delta/t_0)(U+\Delta)/(U-\Delta)$
is the upper limit.
Finally, (iv) EDAP has to be performed
within the coherence time of the system, which is, at low temperatures,
likely in the nanosecond time scale \cite{Petta2004:PRL, Petta2004:P}. Considering full coherence,
the time limitations on the pulse are $\tau_L \alt \tau \alt \tau_{U}$, 
which for our model parameters is 100 ps to 10 ns. 
Since the lower limit is given by energy $t_0^2/\Delta$ which is on the
order of the exchange coupling ($J=t_0^2/U$ in the Hubbard model) 
for our case of $\Delta \approx U$,
the times are similar to those used for spin-based 
quantum computing \cite{Burkard1999:PRB}. The upper limit
$\tau_{U}$ increases with decreasing $|U-\Delta|$. The scheme
will perform quadratically faster for larger couplings.

\begin{figure}
\centerline{\psfig{file=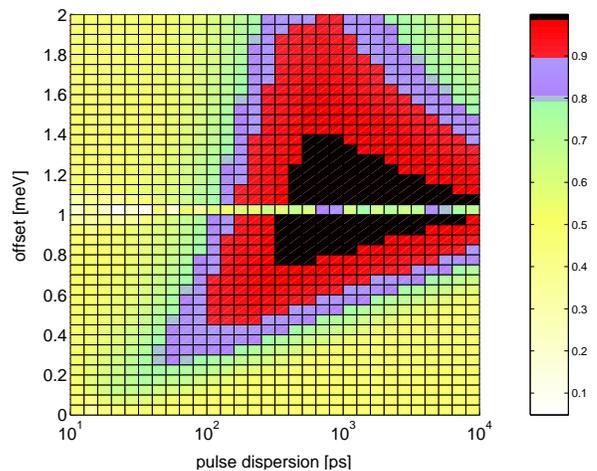,width=0.9\linewidth}}
\caption{Calculated efficiency $w$ as a function of the
pulse dispersion $\tau$ and offset $\Delta$, for the counterintuitive
pulse sequence. The darkest window is for efficiency higher than 98\%,
while the second darkest (red) one is for $w>90\%$.
}
\label{fig:4}
\end{figure}

To identify numerically the regime of applicability of the scheme, 
we define EDAP efficiency $w$ as
\begin{equation}
w=|\langle \Psi(\infty)|S\rangle_{12}|^2 + |\langle \Psi(\infty)|
T_0\rangle_{23} |^2, 
\end{equation}
for a state $\Psi(t)$ with the initial condition
$\Psi(0) = a_{1\uparrow}^+ a_{2\downarrow}^+ |0\rangle$.
This definition is insensitive to the relative phase change, 
and to the relative population of the two states.  
The efficiency is plotted in Fig. \ref{fig:4} as a function of 
$\Delta$ and $\tau$ for the counterintuitive sequence (intuitive
shows the same picture except at $\Delta$ close to 0). The range
of applicability, from 100 ps to 10 ns agrees with our analytical
estimates for our parameters. The graph also shows the predicted
increase of applicable $\tau$ with decreasing $|U-\Delta|$. 
It is evident that our scheme is
very robust, covering large range of spectral values and pulse times. 
The horizontal ``cut'' at $\Delta=U=1$ meV indicates the resonance
oscillations of the $|S\rangle_{12}-|D\rangle_{1}$ pair for which
the efficiency depends on the area of the pulse, $\alpha$, and
thus on $\tau$. The lower limit on $\tau$ can be further reduced by about
a decade (to 50 ps for 98\% efficiency for our parameters) by 
decreasing the delay between the pulses (not shown here).

Efficiency $w$ can be measured
by performing EDAP twice: if the first (distillation) passage results in, say,
singlet, the second (detection) passage should give absence
of charge on dot 3, if $w\approx 1$. 
Another interesting application of EDAP can be in quantifying
the influence of a charge probe on the charge itself. Say, use EDAP to transport 
triplets via $\Psi_1$. Since $n_2 \Psi_1 = \Psi_1$ at all times
(in fact, $\Psi_1$ through $\Psi_3$ are the only eigenstates
of $H(t)$ that are also eigenstates of $n_2$), a measurement of
population on dot 2 should not disturb the state. EDAP efficiency loss
is a measure of the invasiveness of the probe.

In conclusion, we have proposed a robust and realistic all-electronic scheme
for distilling and detecting two-electron spin entanglement in coupled 
quantum dots. The scheme converts entanglement to charge, with close
to 100\% efficiency, by spatially separating
singlet and triplet states within a single superposition. Because the 
entanglement detection is nondestructive the scheme can be used repetitively
to transport entangled pairs as well as to detect disentanglement.  
Although we have used the example of quantum dots, we believe that EDAP 
is general enough to be applicable in other 
physical implementations of quantum information processing.

This work was supported by the US ONR and FWF.

\bibliography{references} 

\end{document}